\documentclass[aps,pra,twocolumn,showpacs,amsmath,superscriptaddress]{revtex4}
\usepackage{graphicx}

\usepackage{bm,amssymb}

\usepackage{graphicx}
\usepackage{dcolumn}
\usepackage{color}
\usepackage{epstopdf}

\begin{document}

\newcommand{\tr}{\mathop{\mathrm{tr}}\nolimits}
\renewcommand{\Re}{\mathop{\mathrm{Re}}}
\renewcommand{\Im}{\mathop{\mathrm{Im}}}


\def\cH{{\cal H}}
\def\cS{{\cal S}}

\newcommand{\beq}{\begin{equation}}
\newcommand{\eeq}{\end{equation}}
\newcommand{\barr}{\begin{eqnarray}}
\newcommand{\earr}{\end{eqnarray}}
\newcommand{\andy}[1]{[#1]}

\newcommand{\REV}[1]{\textbf{\color{red}#1}}
\newcommand{\BLUE}[1]{\textbf{\color{blue}#1}}
\newcommand{\GREEN}[1]{\textbf{\color{green}#1}}

\title{Hausdorff clustering }
\author{Nicolas Basalto}
\affiliation{Market Risk Management, Unicredito Italiano, Milano, Italy}
\author{Roberto Bellotti}
\affiliation{Dipartimento di
Fisica, Universit\`a di Bari, I-70126 Bari, Italy}
\affiliation{Istituto Nazionale di Fisica Nucleare, Sezione di Bari,
I-70126 Bari, Italy}
\affiliation{TIRES, Center of Innovative Technologies for Signal
Detection and Processing, Bari, Italy.}
\author{Francesco De Carlo}
\affiliation{Dipartimento di
Fisica, Universit\`a di Bari, I-70126 Bari, Italy}
\affiliation{Istituto Nazionale di Fisica Nucleare, Sezione di Bari,
I-70126 Bari, Italy}
\author{Paolo Facchi}
\affiliation{Dipartimento di
Matematica, Universit\`a di Bari, I-70125 Bari, Italy.}
\affiliation{Istituto Nazionale di Fisica Nucleare, Sezione di Bari,
I-70126 Bari, Italy}
\author{Ester Pantaleo}
\email{ester.pantaleo@ba.infn.it}  \affiliation{Dipartimento di
Fisica, Universit\`a di Bari, I-70126 Bari, Italy}
\affiliation{Istituto Nazionale di Fisica Nucleare, Sezione di Bari,
I-70126 Bari, Italy}
\author{Saverio Pascazio}
 \affiliation{Dipartimento di
Fisica, Universit\`a di Bari, I-70126 Bari, Italy}
\affiliation{Istituto Nazionale di Fisica Nucleare, Sezione di Bari,
I-70126 Bari, Italy}

\begin{abstract}
A clustering algorithm based on the Hausdorff distance is introduced
and compared to the single and complete linkage. The three
clustering procedures are applied to a toy example and to the time
series of financial data. The dendrograms are scrutinized and their
features confronted. The Hausdorff linkage relies of firm
mathematicl grounds and turns out to be very effective when one has
to discriminate among complex structures.
\end{abstract}

\maketitle

\section{Introduction}
\label{sec:intro}

Clustering is the classification of objects into different groups
according to their degree of \emph{similarity} \cite{fukunaga}. A number
of criteria can be used to define this intuitive (and central)
concept, leading in general to different partitions. Due to this
arbitrariness, clustering is an inherently ill-posed problem, as a
given data set can be partitioned in many different ways without any
particular reason to prefer one solution to another. It is clear
that a clustering technique can be profoundly influenced by the
strategy adopted by the observer and his/her own ideas and
preconceptions on the problem.

Clustering algorithms can be classified in different ways
according to the criteria used to implement them \cite{jain0}:

(A)
If, for example, one focuses on the solution, a fundamental
distinction can be drawn between
\textit{hierarchical} and \textit{partitive} techniques.
Hierarchical methods yield nested partitions, in which any cluster
can be further divided in order to observe its underlying structure.
Typical examples are the \textit{agglomerative} and
\textit{divisive} algorithms that produce \textit{dendrograms}
\cite{jain}. On the other hand, partitional methods provide only one
definite partition which cannot be analyzed in further details.

(B)
By contrast, if one focuses on data representation, two schemes are
possible: \textit{central} \cite{central} and \textit{pairwise}
\cite{duda,hofmann} clustering. In central clustering, the data are
described by their explicit coordinates in the feature space and
each cluster is represented by a \textit{prototype} (for instance,
the mean vector and the corresponding spread). In pairwise
clustering, the data are indirectly represented by a
\textit{dissimilarity} matrix, which provides the pairwise
comparison between different elements. Clearly, the choice of the
measure of dissimilarity is not unique and the performance of any
pairwise method strongly depends on it.

(C)
Finally, if one focuses
on the strategy of the algorithm, two approaches can be adopted:
\textit{parametric} and \textit{non-parametric} clustering.
Parametric algorithms are adopted when some \emph{a priori}
knowledge about the clusters is available and this information is
used to make some assumptions on the underlying structure of the
data.
\emph{Vice versa}, the non-parametric approach to clustering may
represent the optimal strategy when there is no prior knowledge
about the data. In general, these methods follow some local
criterion for the construction of the clusters, such as, for
instance, the identification of high density regions in the data
space
\cite{fukunaga}.

From the mathematical point of view, given a set of objects
$\mathcal{S}\equiv\{s\}$, an allocation function $m:
s\in\mathcal{S}\rightarrow \{1,2,\dots,k\}$, must be defined so that
$m(s)$ is the class label and \textit{k} the total number of
clusters (which we assume to be finite for simplicity);
\textit{k} may be chosen \textit{a priori} or computed within the
algorithm. The aim of a clustering procedure is to select, among all
possible allocation functions, the one performing the best partition
of the set ${\cal S}$ into subsets $\mathcal{G}_{\alpha}\equiv\{ s
\in \mathcal{S} : m(s) = \alpha\}$ \; $(\alpha=1,\ldots,k)$,
relying on some measure of similarity. The space of any clustering
solution is the set $\mathcal{M}$ of all possible allocation functions.

In this article we will focus on a class of clustering techniques
called
\emph{linkage algorithms}. Linkage algorithms are hierarchical,
agglomerative and non-parametric methods that merge, at each step,
the two clusters with the smallest dissimilarity, starting from
clusters made of a single element, ending up in one cluster
collecting all data. We will analyze the so-called single and
complete linkage methods and will introduce a linkage method based
on Hausdorff's distance. We will use  as a mathematical definition
of dissimilarity a suitable metric in the space of the partitions of
the given data set \cite{BBDFPP}. Notice that in general a
similarity measure need not be a distance in the mathematical sense;
on the other hand, if one aims at clustering in a parameter space, a
distance could be the best choice because it does not introduce any
degree of arbitrariness. It is worth stressing that alternative
philosophies are also possible, in which the clustering algorithm is
governed by purely topological notions and unveils efficient
collective dynamics in animal behavior \cite{parisi}. A comparison
among these methods belongs to the realm of statistical mechanics
and is beyond the scope of this article. See \cite{virasoro} for an
excellent discussion.

We will focus on finite sets and clusters, although we will keep our
analysis on the metric features of the relevant spaces as general as
possible. We will start in Sec.\ \ref{sec:Preliminares} by reviewing
and clarifying some mathematical concepts concerning distance and
linkage methods, focusing on the single and complete linkage
algorithms in Sec.\ \ref{sec:Complete_Single}. The Hausdorff
distance and the related clustering procedure will be introduced in
Sec.\ \ref{sec:Hausdorff}. Section \ref{sec:Test} is devoted to the
comparison of the different methods on some data sets, including
both a toy problem and a case study on financial time series. Some
conclusions are drawn in Sec. \ref{sec:summa}.

\section{Preliminares}
\label{sec:Preliminares}

\subsection{Distances and pseudodistances}
\label{sec:Distances}

We start by recalling the mathematical definition of distance. Given
a set ${\cal S}$, a distance (or a metric) $\delta$ is a
non-negative application
\beq
\delta: {\cal S} \times {\cal S}
\longrightarrow \mathbb{R}_+  \label{distfunc}
\eeq
on $\mathbb{R}_+=[0,\infty)$, endowed with the following properties,
valid $\forall x,y \in \cS$:
\barr
& & \delta(x,y)=0 \quad \Longleftrightarrow \quad x=y, \label{assio_1} \\
& & \delta(x,y)=\delta(y,x), \label{assio_2} \\
& & \delta(x,y)\leq \delta(x,z)+\delta(y,z) , \quad \forall z \in
{\cal S} \label{assio_3}
\earr
Incidentally, notice that symmetry (\ref{assio_2}), as well as
non-negativity, are not independent assumptions, but easily follow
from (\ref{assio_1}) and the triangular inequality (\ref{assio_3}).
If the triangular inequality is written as
\beq
\delta(x,y)\leq \delta(x,z)+\delta(z,y) , \quad \forall z \in
{\cal S} , \label{assio_4}
\eeq
as is often the case, symmetry (\ref{assio_2}) must be independently
postulated. We will henceforth denote a metric space by
$(\mathcal{S},\delta)$.

An application (\ref{distfunc}) is a pseudometric \cite{barile} if
property (\ref{assio_1}) is weakened:
\beq x=y \quad \Longrightarrow \quad
\delta(x,y)=0~\label{assio1_min1}.
\eeq
In such a case, \emph{distinct} elements of the set ${\cal S}$ can
be at a null distance. A set endowed with a (pseudo)metric is called
a (pseudo)metric space.

\subsection{Linkage algorithms}
\label{sec:Linkage}

Linkage algorithms are hierarchical methods, yielding a clustering
structure that is usually displayed in the form of a tree or
dendrogram \cite{jain}. We will adopt an agglomerative algorithm,
where the clusters are linked through an iterative process, whose
successive steps are the following. Given a data set ${\cal S}$,
made up of $n$ elements, at the first level (leaves of the
dendrogram) the number of classes is equal to the number of
elements. We assume (without loss of generality) that ${\cal S}$ is
a metric space \footnote{A metric can always be introduced, in any
(finite or infinite) set. The issue here is to understand whether
such a metric is ``physically" meaningful. A good choice makes the
difference between a natural clustering procedure and an artificial
one.}. At the first iteration the two closest elements are clustered
together, reducing the number of classes to $n-1$ (if more than two
elements are at the closest distance, we pick a random couple among
them). At the second iteration one has to tackle the subtler problem
of defining a distance between the remaining elements of ${\cal S}$
and the first cluster formed. When this is done, the distances are
recomputed and the two closest objects are joined. At the following
iterations one has to tackle the much more subtle problem of
defining a distance among classes. Clearly, this can be done in a
variety of different ways and entails further elements of
arbitrariness. Assume that this procedure can be carried out
consistently. After $n$ steps, all the points are grouped together
in one cluster, corresponding to the whole data set. The
agglomerative procedure is reversed in a straightforward way in the
so-called divisive approach: starting from one single cluster, this
is iteratively divided into smaller and smaller ones, until single
elements are obtained.

The most commonly used algorithms of this type are the ``complete"
and the ``single" linkage, that differ in the definition of
``distance" between subsets of points. In the next section we will
briefly review these two algorithms.
\begin{figure}[h]
\includegraphics[width=0.47\textwidth]{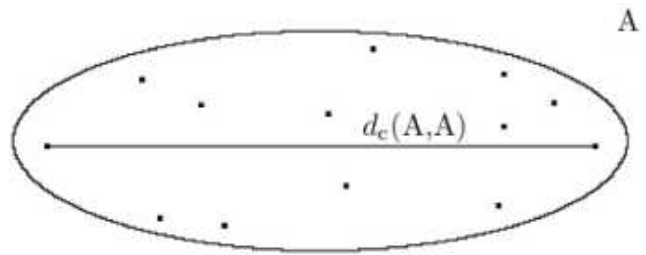}
\caption{For a set $A$ containing more than one element, $d_c(A,A)
\neq 0$ and neither (\ref{assio_1}) nor (\ref{assio1_min1}) are
valid.} \label{fig:dist_max_aa}
\end{figure}

\section{Complete and Single Linkage}
\label{sec:Complete_Single}

\subsection{``Distances"}
\label{sec:nodist}

Linkage algorithms differ from each other for the different
similarity criteria used to build the clusters. An optimal criterium
would rely on a metric $d$ defined on the subsets of the parent
space $\cS$:
\beq d: \mathcal{K}({\cal S}) \times \mathcal{K}({\cal S})
\longrightarrow \mathbb{R}_+ , \label{distPS}
\eeq
where $\mathcal{K}({\cal S})$ is the collection of all the nonempty
compact subsets of $\cS$. (We restrict the metric to the above class
of subsets in order to avoid some patologies, see later.) Such a
metric can be defined in a natural way by using the original metric
$\delta$ defined on $\cS$. If $A$ and $B$ are two non empty compact
subsets of ${\cal S}$, the complete and single linkage ansatzs make
use of the following ``distances"
\beq
d_c(A,B) = \sup_{a\in A, b\in B}\delta(a,b),
\label{dist_max}
\eeq
\beq d_s(A,B) = \inf_{a\in A, b\in B} \delta(a,b) ,
\label{dist_min}
\eeq
respectively. However, it is easy to check that neither one of the
above functions is a \emph{bona fide} distance in the mathematical
sense. The function (\ref{dist_max}) is obviously nonnegative and
symmetric, so (\ref{assio_2}) is valid. Moreover, the triangular
inequality (\ref{assio_3}) is satisfied:
\begin{eqnarray}
d_c(A, B) &=& \sup_{a\in A, b\in B} \delta(a,b) \nonumber \\
& \leq & \sup_{a\in A, b\in B,c\in C}(\delta(a,c)+\delta(b,c)) \nonumber \\
& \leq & \sup_{a\in A, b\in B,c\in C}\delta(a,c)+\sup_{a\in A, b\in B,c\in C}\delta(b,c) \nonumber \\
& = & \sup_{a\in A, c\in C}\delta(a,c)+
\sup_{b\in B,c\in C}\delta(b,c) \nonumber \\
& = & d_c(A,C) + d_c(B,C)~.\label{dim_ass_13}
\end{eqnarray}
Yet, property (\ref{assio_1}) is not valid in general, as for a set
$A$ made up of more than one element, the distance of $A$ from
itself equals the distance between its farthest objects:
\beq
d_c(A,A)\neq 0~. \label{dAAnot0}
\eeq
This is graphically displayed in Fig.\ \ref{fig:dist_max_aa} and
shows that (\ref{dist_max}) is not even a pseudodistance
\footnote{We are excluding the very particular case when
$\delta$ itself is a pseudodistance and all the elements of the
cluster are at a vanishing pseudodistance $\delta$. Notice however
that when one focuses on an iterative clustering algorithm, in all
nontrivial cases, some clusters must eventually acquire a
nonvanishing distance from themselves at some iteration.}.

Intuitively, this is not an important issue for ``small" sets, but
it becomes an increasingly serious problem for ``larger" sets.
Clearly, the notions of ``small" and ``large" must be properly
defined: for a compact metric space of size $R$, we may say that a
subset of size $r$ is small if $r \ll R$ (say by at least one order
of magnitude) \footnote{For non-compact sets whose subsets are
uniformly distributed with linear density $\alpha$, a subset is
``small" if its size $r \ll
\alpha^{-1}$.}. This situation will directly concern us
in the next sections.

\begin{figure}[h]
\includegraphics[width=0.47\textwidth]{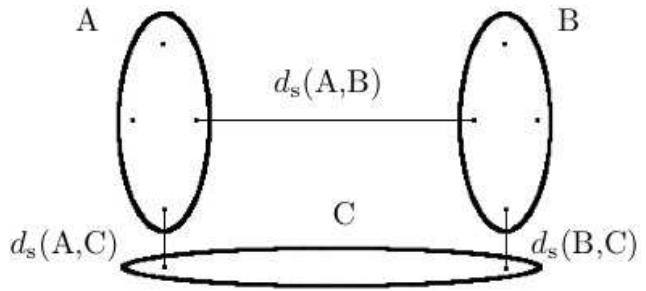}
\caption{Three sets $A$, $B$, $C$, containing each two elements,
for which $d_s$ does not satisfy the triangular inequality:
$d_s(A, B)> d_s(A, C) + d_s(B, C)$.} \label{fig:dist_ABC}
\end{figure}

Consider now the second function (\ref{dist_min}), which is non
negative and symmetric, so (\ref{assio_2}) is valid. Notice that
the pseudometric property is satisfied
\beq A=B  \quad
\Longrightarrow \quad d_s(A,B)=0, \label{assio1_minrep}
\eeq
although the converse is not true [so that property (\ref{assio_1})
is not valid]: consider for instance two sets $A$ and $B$ such that
$A\cap B\neq
\varnothing$: in this case $d_s(A,B)=0$, as this is, by definition,
the distance $\delta$ of a common element from itself.
Finally, the triangular inequality (\ref{assio_3}) is not verified,
as can be easily inferred by looking at the counterexample in Fig.\
\ref{fig:dist_ABC}, for which
\beq
d_s(A, B)> d_s(A, C) + d_s(B,C) .
\eeq
The function $d_s$ is therefore neither a metric nor a pseudometric.
As we shall see in Sec.\ \ref{sec:comments}, this problem gives rise
to the chaining effect.

\subsection{Finite sets}
\label{sec:finsets}

We will look explicitly at the practical case in which
(\ref{dist_max}) and (\ref{dist_min}) are evaluated on
\emph{finite} sets. It is therefore convenient to specialize the
formulas of the preceding section to such a situation. Let
$A=\{a_i\}_{i=1,\ldots,I}$ and $B=\{b_j\}_{j=1,\ldots,J}$ be two
finite sets and \beq \delta_{ij}=\delta(a_i,b_j) \label{dij} \eeq the
distance between any two elements of $A$ and $B$. The
$\delta_{ij}$'s can be arranged in a $I \times J$ ``distance"
matrix. Equations (\ref{dist_max}) and (\ref{dist_min}) read then
\beq d_s(A,B) = \underset{i\in A} \min~\underset{j\in B} \min
~\delta_{ij} , \label{dist_single} \eeq \beq d_c(A,B) = \underset
{i\in A} {\max}~\underset {j\in B} {\max}~\delta_{ij},
\label{dist_complete} \eeq for the single and complete linkage
algorithms, respectively. In practice, this amounts to determine the
smaller and the larger value among the rows and the columns of the
distance matrix, respectively, a task that can be performed in a polynomial time. These formulas will be applied in the
following examples.

\subsection{Comments}
\label{sec:comments}

\begin{figure}[h]
\includegraphics[width=0.47\textwidth]{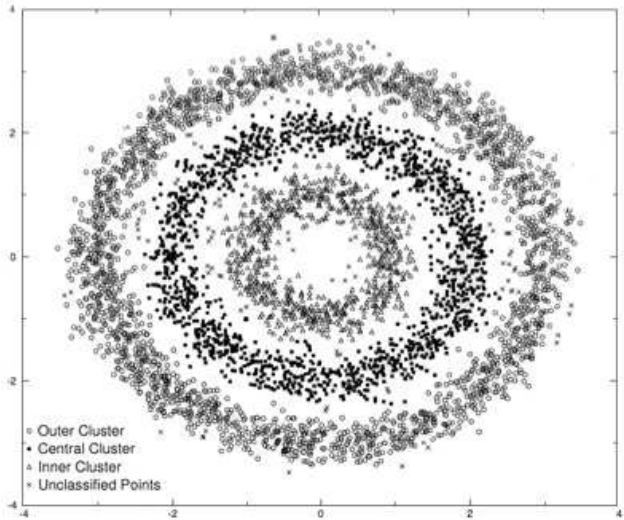}
\caption{Concentric clusters analyzed in terms of the
single linkage algorithm. This procedure is very efficient in
discriminating nested structures of this kind. For the same
reason, it suffers from the so-called ``chaining effect."}
\label{fig:SPC_1}
\end{figure}

It is worth commenting on the features of the two clustering ansatzs
introduced, emphasizing their limits and positive aspects. The
single linkage algorithm tends to yield elongated clusters, which
are sometimes difficult to understand and poorly significant
\cite{jain}: this is known as
\textit{chaining effect}. On the contrary, the complete linkage has
the advantage of clustering ``compact'' groups and produces well
localized classes. In general, the partitions obtained using it are
more significant. Its major disadvantage is that it does not set
equal to zero the distance of a ``compact" set from itself [see Eq.\
(\ref{dAAnot0}) and Fig.\
\ref{fig:dist_max_aa}], performing \emph{de facto} a coarse
graining. In few words, $d_c$ looks at the data points with a
``minimal resolution'' (that is also, unfortunately, cluster
dependent) and is unable to recognize the complexity of a finely
structured cluster and to extract ``nested" clusters, such as those
displayed in Fig.\
\ref{fig:SPC_1} \cite{domany}. Notice that, by contrast, such
``nested" clusters are very efficiently detected by the single
linkage algorithm, as shown in Fig.\
\ref{fig:SPC_1}.

In the next section we shall introduce a procedure that is somewhat
``in between" single and complete linkage and makes use of an
underlying \emph{bona fide} distance. This will have some
advantages, also from a conceptual viewpoint, as it enables one to
rest on firm mathematical background.

\section{Hausdorff distance and Hausdorff linkage}
\label{sec:Hausdorff}

In the light of the discussion of the preceding section, it appears
convenient to approach the clustering problem from a ``neutral"
perspective, by looking for a linkage algorithm based on a
well-defined mathematical similarity criterium. In order to do this,
we will use a distance function introduced by Hausdorff
\cite{hausdorff}.

\subsection{Hausdorff distance}
\label{sec:Hdist}

Given a metric space $(\cS, \delta)$, the distance between a point
$a\in\cS$ and a (nonempty and compact) subset $B \in \mathcal{K}(\cS)$ is naturally given by
\begin{equation}
\tilde d(a; B)=\inf_{b\in B} \delta(a,b)
\end{equation}
Given a
subset $A \in \mathcal{K}(\cS)$, consider the function
\beq
\tilde d(A;B)=\sup_{a\in A} \tilde d(a; B) = \sup_{a\in
A}\inf_{b\in B} \delta(a,b),
\label{openhaus1}
\eeq
that measures the largest distance $\tilde d(a; B)$, with $a\in A$.
Note that here the strategy is opposite to that used with the single
linkage ``distance" (\ref{dist_min}), where one considers instead
the \emph{smallest} distance $\tilde d(a; B)$, with $a\in A$. The
function (\ref{openhaus1})  is not symmetric, $\tilde d(A;B)
\neq\tilde d(B;A)$, and therefore is not a \emph{bona fide}
distance, as it does not satisfy (\ref{assio_2}). The Hausdorff
distance \cite{hausdorff} between two sets $A,B\in\mathcal{K}(\cS)$ is defined as the largest between the two numbers:
\barr
d_{\rm H}(A,B)=\max\{\tilde d(A;B), \tilde d(B;A)\},
\earr
namely,
\barr
d_{\rm H}(A,B)= {\max}\{\sup_{a\in A}\inf_{b\in B}\delta(a,b),
\sup_{b\in B}\inf_{a\in A} \delta(a,b) \},
\label{link_haus}
\earr
that is clearly symmetric and satisfies all axioms
(\ref{assio_1})-(\ref{assio_3}).

\begin{figure}
\begin{center}
\includegraphics[width=0.47\textwidth]{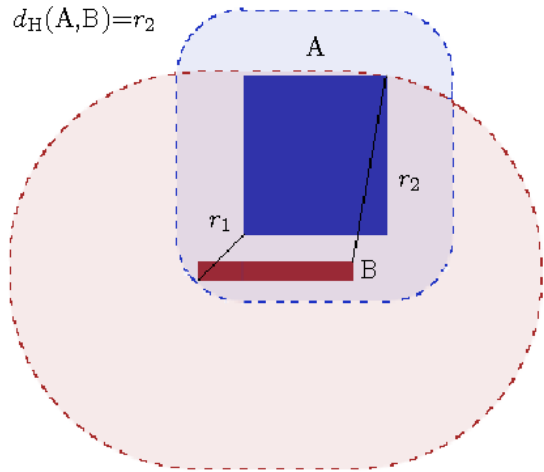}
\end{center}
\caption{Hausdorff distance between two sets $A$ (a square) and $B$
(a rectangle). The open neighborhoods
$N_{r_1}(A)$ and $N_{r_2}(B)$ are shaded, $r_1=\tilde d(B;A)$, $r_2=\tilde d(A;B)$. The Hausdorff distance is $r_{2}$.}
\label{fig:dist_H_AB}
\end{figure}

It is worth discussing a bit more the mathematical features of
$d_{\rm H}$. This will help us grasp its interesting properties,
towards physical applications.

Given a set $A \in\mathcal{K}(\cS)$ and a positive real number $r>0$,
define the {\it open r-neighborhood} of $A$ as:
\beq
N_{r}(A)=\{y:\tilde d(y;A)<r \}~.
\label{openhaus}
\eeq
The Hausdorff distance between two sets $A,B \in\mathcal{K}(\cS)$ can be
reexpressed as
\beq
d_{\rm H}(A,B)={\inf}\{r : A\subseteq N_{r}(B)~\mbox{and}~B\subseteq
N_{r}(A)\}~.\label{dist_haus}
\eeq
Indeed
\begin{eqnarray}
& & d_{\rm H}(A,B) \nonumber\\
& & = {\inf}\{r : A\subseteq N_{r}(B),~B\subseteq
N_{r}(A)\}\nonumber \\
& & ={\inf}(\{r: A\subseteq N_{r}(B)\}\cap
\{r : B\subseteq N_{r}(A)\}) \nonumber\\
& & = {\max}\{{\inf}\{r : A\subseteq
N_{r}(B)\},~{\inf}\{r : B\subseteq
N_{r}(A)\}\}\nonumber\\
\end{eqnarray}
and since
\begin{eqnarray}
{\inf}\{r : A\subseteq N_{r}(B)\} &=& \underset{x\in A}\sup ~
\inf\{r : x\in N_{r}(B)\} \nonumber\\
&=& \underset {x\in A ~~y\in B}{\sup~\inf}~\delta(x,y)~,
\end{eqnarray}
and analogously for ${\inf}\{r : B\subseteq N_{r}(A)\}$, one gets
again (\ref{link_haus}). Stated differently, the Hausdorff distance
can also be defined as the smallest radius $r$ such that $N_{r}(A)$
contains $B$ and at the same time $N_{r}(B)$ contains $A$.

In words, the Hausdorff distance between $A$ and $B$ is the smallest
positive number $r$, such that every point of $A$ is within distance
$r$ of some point of $B$, \emph{and} every point of $B$ is within
distance $r$ of some point of $A$. The geometrical meaning of the
Hausdorff distance is best understood by looking at an example, such
as that in Fig.\
\ref{fig:dist_H_AB}. We emphasize that the Hausdorff metric on the
subsets of $\cS$ is defined in terms of the metric $\delta$ on the
points of $\cS$.

The Hausdorff distance enjoys a number of interesting features, that
are worth discussing. We have defined $d_{\rm H}$  only on nonempty
compact sets for the following reasons. Consider for example the
real line. Then, by adopting the convention ${\inf}\{
\varnothing \}=\infty$ \footnote{This is motivated by thinking of
unbounded sets: the Hausdorff distance between a point $A=\{a\}$ and
a set $B$ having an accumulation point at infinity (such as a
straight line) is indefinitely large, for no open $r$-neighborhood
$N_{r}(A)$ will ever contain $B$, no matter how large $r$.}, one
gets $\forall x, d_{\rm H}(\varnothing,x)=\infty$, which is not
allowed by any definition of metric. This suggests that we should
restrict our attention to nonempty sets. Moreover, $d_{\rm
H}(\{0\},[0,\infty))=\infty$, which is again not allowed. We then
restrict the use of $d_{\rm H}$ only to bounded sets. Finally, the
Hausdorff distance between two not equal sets could vanish [which
would make $d_{\rm H}$ a pseudometric, see  (\ref{assio1_min1})]:
for instance $d_{\rm H}((0,1),[0,1])=0$. Therefore we will restrict
the application of $d_{\rm H}$ only to closed sets.

More generally, it is easy to prove the following \\
\textbf{Theorem}: The Hausdorff function $d_{\rm H}$ is a metric
on the set ${\cal K}(\cS)$. Moreover, if $(\cS,\delta)$ is a
complete metric space, then the space $({\cal K}(\cS), d_{\rm H})$
is also complete.

Although of an abstract nature, this is of
\emph{physical significance}, as it enables one to be confident about
the metric properties of ${\cal K}(\cS)$ even for fine-structured
clusters. Notice that the property of completeness could not even be
conceived for the ``distance" $d_c$ used for the complete linkage in
the last section. In conclusion,
\beq
d_{\rm H}: {\cal K}(\cS) \times {\cal K}(\cS)
\longrightarrow \mathbb{R}_+  \label{Hfunc}
\eeq
is a complete metric. In the cases of interest, $\cS$ will be a complete metric space, e.g.,
an Euclidean space.

We close this section with two remarks. First, if the data set is
finite and consists of $N$ elements, all distances can be arranged
in a $N \times N$ matrix $\delta_{ij}$ and Eq.\ (\ref{link_haus})
reads
\beq
d_{\rm H}(A,B)=\max\{ \max_{i\in A}\min_{j\in B} \delta_{ij}~,
\max_{j\in B}\min_{i\in A} \delta_{ij}\},
\label{link_haus_finite}
\eeq
which is a very handy expression, as it amounts to finding the
minimum distance in each row (column) of the distance matrix, then
the maximum among the minima. The two numbers are finally compared
and the largest one is the Hausdorff distance. This sorting
algorithm is efficient and can be easily implemented.

Second, $\forall A, B\in\mathcal{K}(\cS)$
\beq
d_s (A,B) \leq d_{\rm H}(A,B) \leq d_c (A,B).
\label{HCineq}
\eeq This is a simple consequence of
(\ref{link_haus}) and the definitions (\ref{dist_max}) and (\ref{dist_min}) [or
(\ref{link_haus_finite}), (\ref{dist_complete}) and  (\ref{dist_single}) in the discrete
case]. In some sense, $d_c$ overestimates the distance between two
given sets, essentially because it includes in such a distance the
very ``size" (\ref{dAAnot0}) of the set (see Fig.\
\ref{fig:dist_max_aa}). On the other hand, $d_s$ underestimates it.
As we shall see, this has important
consequences when one clusters complex and/or large sets.

\subsection{Hausdorff linkage}

We shall take the  Hausdorff distance as our dissimilarity measure.
This distance naturally translates in a linkage algorithm: at the
first level each element is a cluster, the Hausdorff distance
between any pair of points reads
\beq
d_{\rm H}(\{i\},\{j\}) = \delta_{ij}
\eeq
and coincides with the underlying metric. The two elements of
$\cS$ at the shortest distance are then joined together in a
single cluster. The Hausdorff distance matrix is recomputed,
considering the two joined elements as a single set. This
iterative process goes on until all points belong to a single
final cluster.

Clearly, when evaluating distances among single elements (points),
the three procedures $d_{\rm H}$, $d_s$, $d_c$ yield the same
result. The output of the single linkage algorithm will clearly
differ very quickly from the other two, due to the drawbacks of the
chaining effect. On the other hand, the differences between
Hausdorff and complete linkage will become apparent only later in
the clustering process. This is a consequence of the fact that the
functions $d_{\rm H}$ and $d_c$ yield the same value when evaluated
on a single element $\{a\}$ and a composite set $B$. Indeed, from
(\ref{link_haus}):
\begin{eqnarray}
& & d_{\rm H}(\{a\},B) \nonumber \\
& & = {\max}\{\underset {x\in\{a\} ~y\in B}{\sup ~\inf}~\delta(x,y),
\underset
{~~y\in B ~x\in\{a\}}{\sup ~\inf}\delta(x,y)\} \nonumber \\
& & =  {\max}\{\underset {y\in B}{\inf}~\delta(a,y),~\underset {y\in
B}{\sup}~\delta(a,y)\} \nonumber \\
& & =  \underset {y\in B}{\sup}~\delta(a,y)
\nonumber \\
& & =  d_c(\{a\},B)~.
\end{eqnarray}
As a consequence of this property, at the lowest levels the
Hausdorff linkage will yield a partition that is very similar to
that obtained by the complete linkage algorithm. As the clustering
procedure goes on, the two methods will differ from each other,
because of their different criteria in evaluating distances, leading
to different aggregations of more complex classes. It is at this
point that the output of the complete linkage becomes less reliable,
as a consequence of (\ref{dAAnot0}) and (\ref{HCineq}). As discussed
after Eq.\ (\ref{dAAnot0}), we expect this problem to become serious
for ``large" sets, of size comparable to that of the parent space.

The partitions obtained by the Haudorff linkage algorithm will be intermediate between
those obtained by the other two procedures. We shall now compare
the three clustering methods, first on an artificial set of points
in a two dimensional Euclidean space, then on financial time
series.

A final comment is in order. Given a distance matrix, any clustering
procedure will yield a tree and an ultrametric, entailing a loss of
information on the data set. However, this appears necessary and is
inherent in any clustering procedure.

\section{Applications}
\label{sec:Test}

\subsection{Two-dimensional data set}
\label{sec:occhiali}

\begin{figure}[h]
\includegraphics[width=0.47\textwidth]{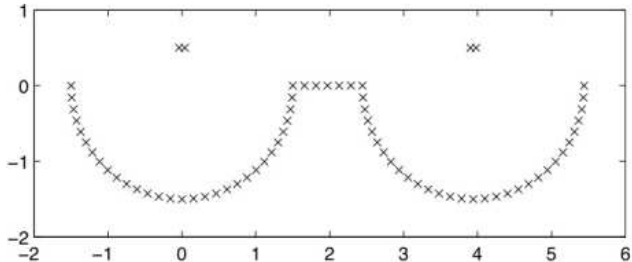}
\caption{A two-dimensional toy sample: a pair of ``glasses" (each one made up of 31
points) connected by a short horizontal ``bar" (5 points) and two
``pupils" (each one made up of 2 points), for a total of $n = 71$
points.}
\label{fig:set}
\end{figure}

Let us analyze the effect of the single, complete and Hausdorff
linkage algorithms on the data set shown in Fig.\ \ref{fig:set}. This
is a discrete set of points in the plane, resembling a pair of
``glasses" (each one made up of 31 points) connected by a short
horizontal ``bar" (5 points) and two ``pupils" (each one made up of
2 points), for a total of $n = 71$ points.

\begin{figure*}
\includegraphics[width=0.95\textwidth]{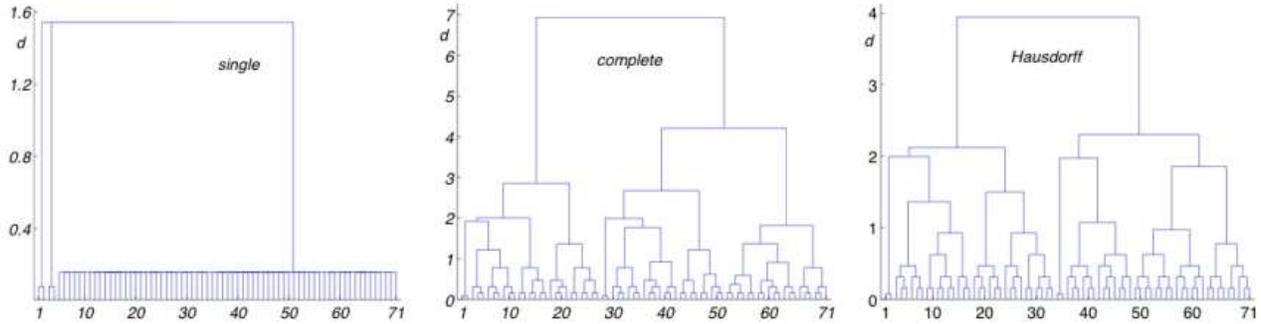}
\caption{Dendrograms generated by the
single, complete and Hausdorff linkage, for the data set of Fig.\
\ref{fig:set}. }
\label{fig:dendrocchiali}
\end{figure*}

This example aims at showing how difficult it can be to discriminate
between complete and Hausdorff linkage: while the single linkage
will obviously suffer from the chaining effect (and will cluster
points at the opposite sides of the figure), the other two
procedures will perform in a similar fashion at the beginning,
yielding different clusters only when the classes become more
complex.

The dendrograms generated by the three algorithms are shown in Fig.\
\ref{fig:dendrocchiali}. The chaining effect of the single linkage
is apparent. This can be an advantage if one wants to bring to light
the presence of a ``continuous" line of points; it is a drawback in
a parameter space because data characterized by opposite values of
the parameter on the abscissa in Fig.\ \ref{fig:set} are clustered
together. As anticipated, a discrimination between the two other
methods is more difficult. However, as discussed after Eq.\
(\ref{dAAnot0}), the differences should become apparent for ``large"
sets, of size comparable to that of the parent space: for a parent
space made up of $n=71$ (approximately linearly distributed) points,
we expect this effect to show up for sets made up of more than 7
points, as one can see in Fig.\ \ref{fig:dendrocchiali}.

\begin{figure}[h]
\includegraphics[width=0.47\textwidth]{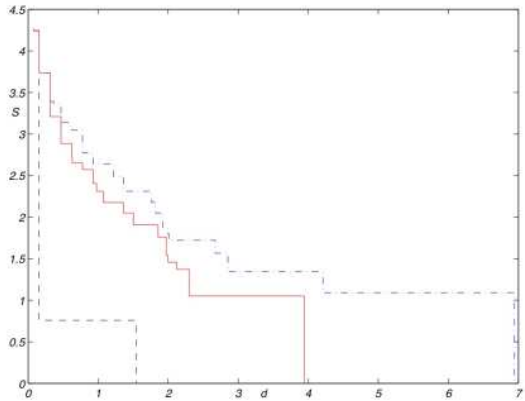}
\caption{Cluster entropies of the dendrograms of Fig.\
\ref{fig:dendrocchiali}.
Dashed black line: single; continuous red line: Hausdorff; blue
dot-dashed line: complete.} \label{fig:entropies}
\end{figure}

\begin{figure}
\includegraphics[width=0.47\textwidth]{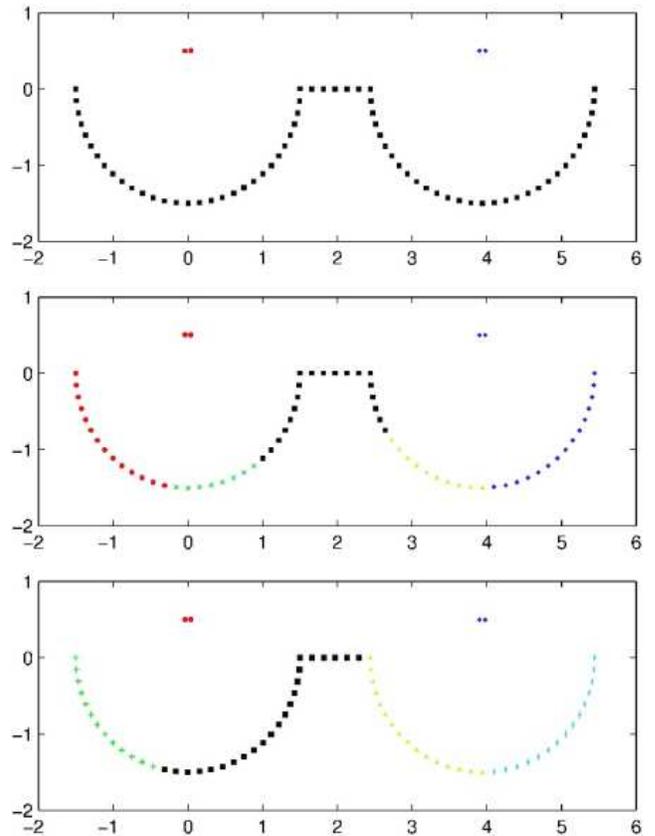}
\caption{Clustering results for single (up), complete
(middle), and Hausdorff (bottom) linkage. Objects belonging to the
same cluster share the same symbol: for example the complete
algorithm groups the ``pupil" on the left (red full circles) with 14
point belonging to its nearest ``glass" (red full circles).}
\label{fig:clusters}
\end{figure}

A proper way to cut the dendrograms could be to search for a stable
partition among the whole hierarchy yielded by the algorithms, in
correspondence to an approximately constant value of the cluster
entropy in a certain range of the dissimilarity measure
\textit{d}
\cite{kaneko}
\begin{equation}
    S(d) = -\sum_{k=1}^{N_{d}}P_{d}(k)\ln P_{d}(k) ~, \label{Boltzmann-entropy}
\end{equation}
where $P_{d}(k)$ is the fraction of elements belonging to cluster
$k$, and $N_{d}$ the number of clusters at level $d$ in the
dendrogram. The complete and Hausdorff entropies corresponding to
the dendrograms in Fig.\ \ref{fig:dendrocchiali} are shown in Fig.\
\ref{fig:entropies}.
We emphasize that, for the case at hand, the data set was
intentionally chosen so that one cannot expect an obvious partition
into ``sensible" clusters. For this very reason, the entropies in
Fig.\ \ref{fig:entropies} display no ``plateau." The optimal cut is
then chosen according to a visual optimization of the clustering
solution. Figure \ref{fig:clusters} shows the selected partitions:
while the single linkage yields a clear chaining effect, both
complete and Hausdorff methods share the positive aspect of
clustering rather ``compact" sets. Moreover, all other clusters
being roughly similar, the Hausdorff procedure is also able to
discriminate the two-points ``pupils" in Fig.\
\ref{fig:set}: in this respect it enjoys the positive spin-offs of
the single linkage algorithm. On the other hand, the complete
linkage algorithm clusters each ``pupil" together with a part of its
nearest ``glass."

\subsection{Financial Data}
\label{sec:Finance}

The use of clustering algorithms can improve the reliability of a
financial portfolio \cite{mantegna_cluster}. Here we apply the
Hausdorff algorithm to the analysis of financial time series
\cite{BBDFPP}. In particular, we focus on the $N=30$ shares
composing the DJIA index, collecting the daily closure prices of its
stocks for a period of 5 years (1998-2002). The companies of the
DJIA stock market are reported in Appendix
\ref{sec-appA}, together with the corresponding industrial areas.

\begin{figure}[h]
\includegraphics[width=0.49\textwidth]{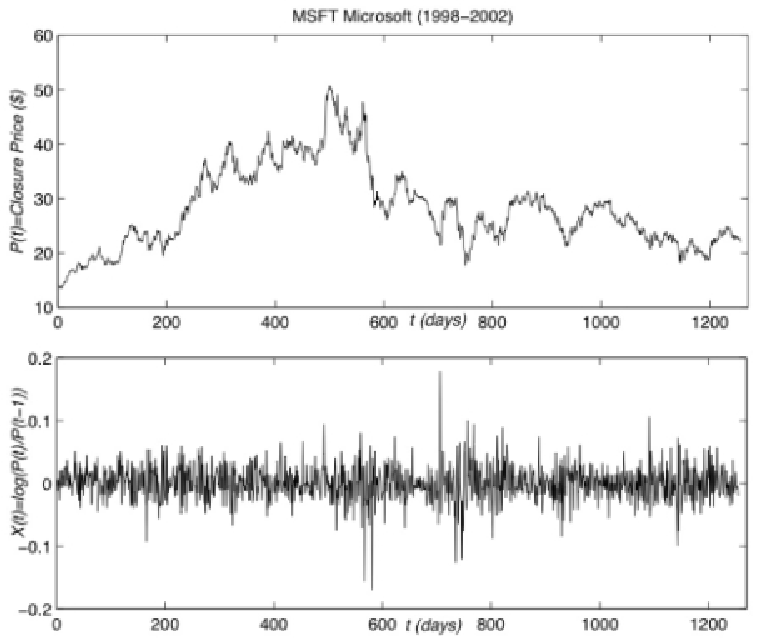}
\caption{Time evolution of the closure price $P(t)$ and the logarithm of the
ratio of consecutive closure prices $X(t)$ [see Eq.\ (\ref{ln_C})]
of a stock value (\textsf{MSFT}), for the period 1998-2002.}
\label{fig:MSFT}
\end{figure}

\begin{figure}[h]
\includegraphics[width=0.3\textwidth]{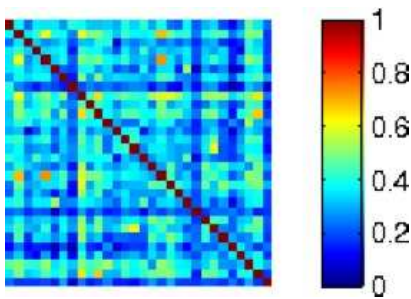}
\caption{Correlation matrix
$\rho(X,Y)$ computed for the year 1998: each element is displayed in
a color scale ranging from blu (minimum value) to red (maximum
value)}
\label{corr98}
\end{figure}


We consider the temporal series of the logarithm of the ratio of two
consecutive closure prices \beq X(t) \equiv \ln \frac{P(t)}{P(t-1)},
\label{ln_C} \eeq where $P(t)$ is the closure price of a
stock at day $t$. Both $P$ and $X$ are very irregular functions of
time, as one can see in Fig. \ref{fig:MSFT}, that displays the
typical behavior of a stock value (\textsf{MSFT}) for the
investigated time period. In order to use the linkage algorithm, we
quantify the degree of similarity between two time series X and Y by
means of the correlation coefficients computed over the investigated
time period:
\beq
\rho(X,Y) =\frac{\mathrm{cov}(X,Y)}{\sigma_X\sigma_Y}= \frac{E[(X-\mu_X)(Y-\mu_Y)]}{\sigma_X\sigma_Y}
\label{rho}
\eeq
where $E$ is the expectation value over the time interval of
interest (one year in our case), $\mu_X=E[X]$ and
$\sigma_X=\sqrt{E[X^2]-\mu_X^2}$. Figure \ref{corr98} shows the
correlation matrix $\rho(X,Y)$ computed for the year 1998: each
element is displayed in a color scale ranging from blu (minimum
value) to red (maximum value). It is worth stressing that almost all
correlation coefficients are positive, with values not too close to
1, thus confirming that, in many cases, stocks belonging to the same
market do not move independently from each other, but rather share a
similar temporal behavior.

\begin{figure}[h]
\includegraphics[width=0.3\textwidth]{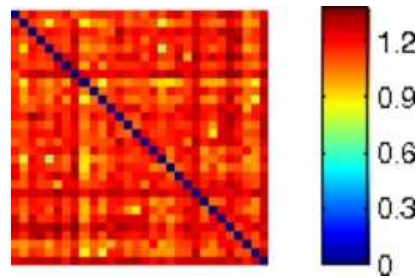}
\caption{Distance matrix $d(X,Y)$ computed for the year 1998: each element is
displayed in a color scale ranging from blu ($d=0$) to red
($d=\sqrt{2}$)}
\label{dist98}
\end{figure}

The metric function we adopted to quantify the time synchronicity
between two stocks is the following
\cite{mantegna,mant_stan,Grilli1}:
\beq
d(X,Y) = \sqrt{2(1-\rho(X,Y))}\label{dist}~.
\eeq
The distance (\ref{dist}) is a proper metric in the parent space,
ranging from 0 for perfectly correlated series $[\rho(X,Y) = +1]$ to
2 for anticorrelated stocks $[\rho(X,Y) = -1]$. The representative
points lie  on a hypersphere and $d(X,Y)$ measures the Euclidean
(and not the geodesic) distance between $X$ and $Y$. Figure
\ref{dist98} shows the distance matrix $d(X,Y)$ computed for the
year 1998: each element is displayed in a color scale ranging from
blu ($d=0$) to red ($d=\sqrt{2}$). The tree structure obtained for
this set was already scrutinized and discussed in Ref.\
\cite{BBDFPP}. We shall focus here on the features of the
dendrograms.

\begin{figure*}
\includegraphics[width=0.95\textwidth]{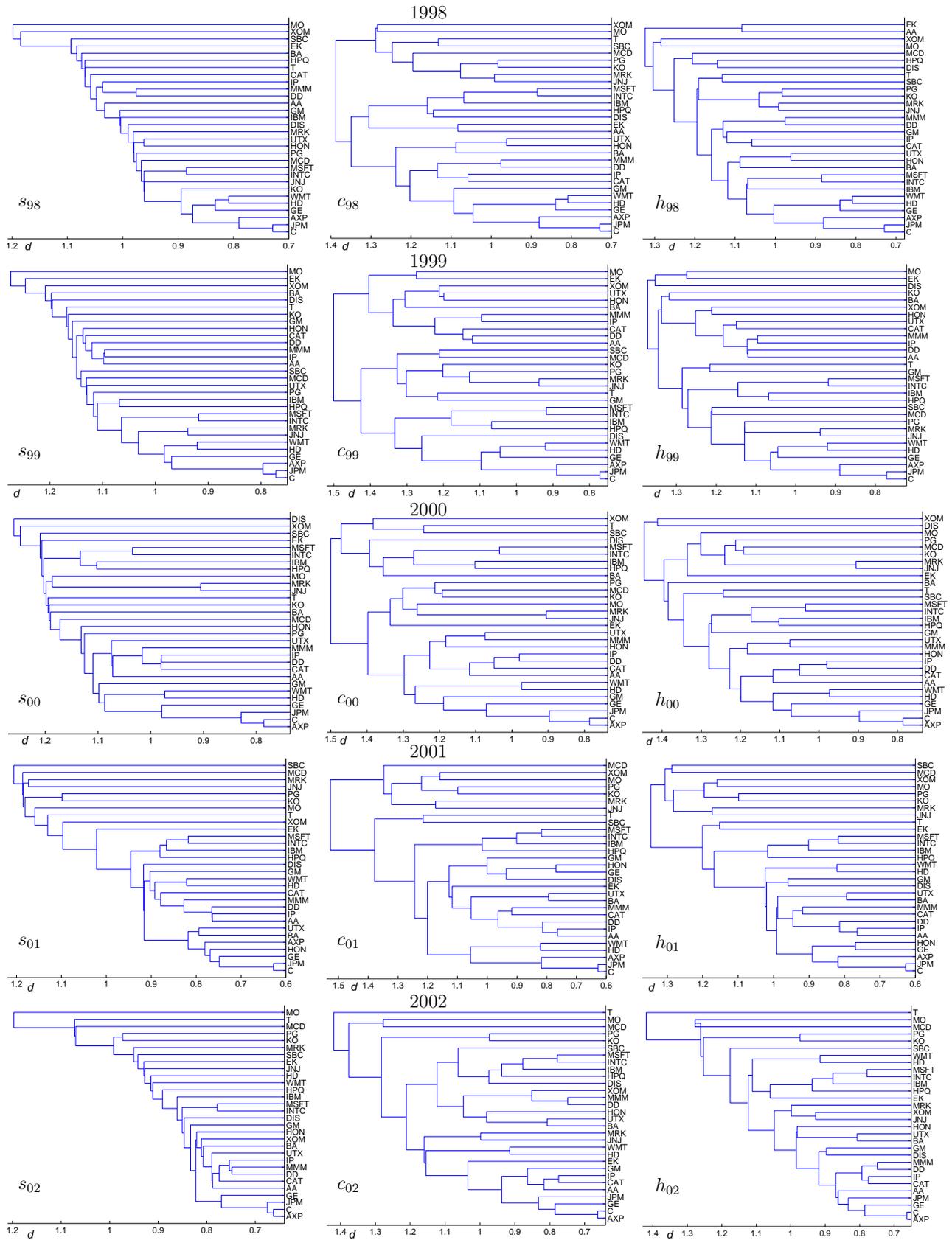}
\caption{Dendrograms obtained by clustering the stocks from 1998 to 2002 for: single
linkage (from $s_{98}$ to $s_{02}$), complete linkage (from $c_{98}$
to $c_{02}$) and Hausdorff linkage (from $h_{98}$ to $h_{02}$). The
acronyms are explained in Appendix \ref{sec-appA}. Some ``backsteps"
can be clearly observed in $h_{02}$. A mathematical explanation of
this phenomenon is given in Appendix
\ref{sec-appB}.
}\label{fig:dendrodj}
\end{figure*}

Figure \ref{fig:dendrodj} shows the dendrograms obtained by
clustering the stocks yearly from 1998 to 2002, with the single,
complete and Hausdorff linkage. Some considerations are in order. As
expected, the single linkage algorithm suffers from the chaining
effect \cite{jain}, which yields elongated clusters: different
points merge into a large cluster almost one at time during the
iterative procedure, with the result of obtaining a poorly defined
tree structure, as it can be clearly observed in Fig.\
\ref{fig:dendrodj} (from $s_{98}$ to $s_{02}$). Wherever one would
choose to cut the dendrogram, no meaningful partition would emerge
out of the hierarchical tree. On the other hand, the dendrograms
obtained by means of both the complete and Hausdorff algorithms show
clear inner structures, corresponding to the branches of the
hierarchical tree. One recognizes the clusters corresponding to
homogeneous (from the industrial viewpoint) groups of companies,
belonging to the same industrial area: this is the case of the money
center banks \{\textsf{C}, \textsf{JPM}\, \textsf{AXP}\}, retail
companies
\{\textsf{HD}, \textsf{WMT}\}, companies dealing with basic
materials \{\textsf{AA}, \textsf{IP}, \textsf{DD}\},  and the
technological core \{\textsf{IBM}, \textsf{INTC}, \textsf{MSFT}\}.

The classification of stocks in terms of their economic homogeneity
as well as the presence of superclusters and homogeneous subgroups
was already discussed in \cite{BBDFPP} and will not be analyzed
here. However, there are characteristic features of the dendrograms
that deserve additional attention. An interesting phenomenon,
consisting in ``backsteps" in the dendrograms, sometimes appears in
the Hausdorff clustering, as shown in $h_{02}$ of Fig.\
\ref{fig:dendrodj}, the dendrogram obtained by clustering the
financial time series in 2002. This pattern is mathematically
spelled out in Appendix \ref{sec-appB}, where its significance is
elucidated in terms of an elementary example (see Fig.\
\ref{fig:backstep}). We take this phenomenon as an indicator of
the potentialities of a clustering algorithm based on the Hausdorff
distance, that could be exploited in a non-hierarchical algorithm,
allowing backsteps and hierarchy breaking.

\section{Summary}
\label{sec:summa}
Clustering is a common practice in the analysis of complex data and
reflects a human compulsion towards classifying objects or physical
phenomena. This can be a difficult task when the phenomena are
complicated and the underlying correlations difficult to bring to
light. We have introduced and analyzed a clustering procedure based
on a \emph{bona fide} distance introduced by Hausdorff. The method,
that relies on an underlying distance among the elements that make
up the ``parent" set, has been compared with both the single and
complete linkage procedures, which only rely on an underlying
dissimilarity measure (not a distance). We first looked at a toy
problem, in which the Hausdorff method has evident advantages in
comparison with the other ones. We then clustered the financial time
series of the DJIA stock market, observing the formation of clusters
of ``homogeneous" companies: the results obtained are significant
from an economical point of view.

An important application of the method introduced here is certainly
in portfolio optimization
\cite{elton,laloux,onnela2,mantegna_cluster,onnela}, where the key issue is
to select one (or a few) stocks that are representative of a given
cluster, characterized by economic homogeneity, reducing maintenance
costs and optimizing risk. Among the possible future developments,
one should test the stability of the method against noise effects
\cite{aste,tumminello} and endeavor to understand the practical
consequences of hierarchy breaking due to the backsteps discussed in
the previous section.

\appendix
\section{Dow Jones stock market companies} \label{sec-appA}
\begin{description}
    \item[AA:] Alcoa Inc. - Basic Materials
    \item[AXP:] American Express Co. - Financial
    \item[BA:] Boeing - Capital Goods
    \item[C:] Citigroup - Financial
    \item[CAT:] Caterpillar - Capital Goods
    \item[DD:] DuPont - Basic Materials
    \item[DIS:] Walt Disney - Services
    \item[EK:] Eastman Kodak - Consumer Cyclical
    \item[GE:] General Electrics - Conglomerates
    \item[GM:] General Motors - Consumer Cyclical
    \item[HD:] Home Depot - Services
    \item[HON:] Honeywell International - Capital Goods
    \item[HPQ:] Hewlett-Packard - Technology
    \item[IBM:] International Business Machine - Technology
    \item[INTC:] Intel Corporation - Technology
    \item[IP:] International Paper - Basic Materials
    \item[JNJ:] Johnson \& Johnson - Healthcare
    \item[JPM:] JP Morgan Chase - Financial
    \item[KO:] Coca Cola Inc. - Consumer Non-Cyclical
    \item[MCD:] McDonalds Corp. - Services
    \item[MMM:] Minnesota Mining - Conglomerates
    \item[MO:] Philip Morris - Consumer Non-Cyclical
    \item[MRK:] Merck \& Co. - Healthcare
    \item[MSFT:] Microsoft - Technology
    \item[PG:] Procter \& Gamble - Consumer Non-Cyclical
    \item[SBC:] SBC Communications - Services
    \item[T:] AT\&T Gamble - Services
    \item[UTX:] United Technology - Conglomerates
    \item[WMT:] Wal-Mart Stores - Services
    \item[XOM:] Exxon Mobil - Energy
\end{description}

\section{}
\label{sec-appB}

\begin{figure}[h]
\includegraphics[width=0.47\textwidth]{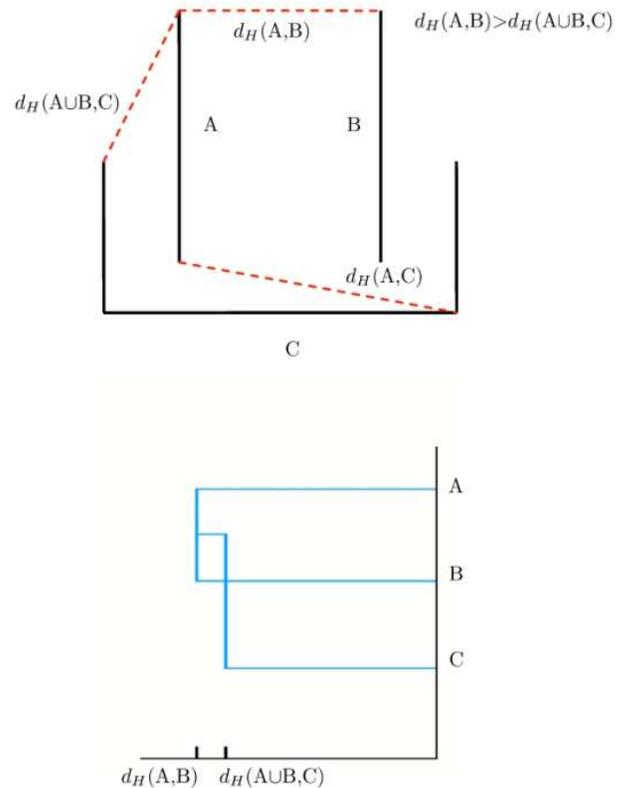}
\caption{Example of a backstep in the Hausdorff
linkage. Given three sets $A$ (a segment), $B$ (another segment) and
$C$ (a ``U") , the Hausdorff linkage algorithm links $A$ and $B$ at
a distance $d_{\rm H}(A,B)$, then links $A\cup B$ and $C$ at a
distance $d_{\rm H}(A\cup B,C)< d_{\rm H}(A,B)$. The set $C$
is nearer to $A\cup B$ than it is to $A$ and $B$ separately. The
corresponding dendrogram is drawn below.}
\label{fig:backstep}
\end{figure}

We explain here the phenomenon of the
backsteps observed in the Hausdorff dendrogram of Fig.\
\ref{fig:dendrodj} (see panel $h_{02}$) and argue that the
Hausdorff hierachical clustering does not exploit all the
potentialities of the Hausdorff distance.

Let us consider the three compact sets of the Euclidean plane shown in Fig.~\ref{fig:backstep}.
Set $A$ is a segment, $B$ is another segment and
$C$ is a polygonal ``U". They are arranged in such a way that
\begin{equation}
d_{\rm H}(A,B) < d_{\rm H}(A,C), \quad \mathrm{and}\quad
d_{\rm H}(A,B) < d_{\rm H}(B,C).
\end{equation}
Therefore, the Hausdorff linkage algorithm starts off by linking $A$
and $B$ at a distance $d_{\rm H}(A,B)$ into a cluster $D=A\cup B$.
But now it happens that the Hausdorff distance between $C$ and
cluster $D$ is smaller than the Hausdorff distance between $A$ and
$B$, namely
\begin{equation}
d_{\rm H}(D,C)=d_{\rm H}(A\cup B,C)< d_{\rm H}(A,B).
\label{eq:diseq}
\end{equation}
Therefore, the set $C$ is nearer to  $D=A\cup B$
than it is to $A$ and $B$ separately,
\begin{equation}
\label{eq:diseq1}
d_{\rm H}(A\cup B,C)< d_{\rm H}(A,C), \;  d_{\rm H}(B,C),
\end{equation}
 and the
corresponding dendrogram exhibits a backstep.

It can therefore happen that two sets, after their aggregation,
become Hausdorff-closer to a third set than they were separately.
This explains (from a mathematical viewpoint) the phenomenon of the
backsteps observed in Fig.\ \ref{fig:dendrodj} (see panel
$h_{02})$.

Therefore, backsteps are a direct consequence of the very definition
of the Hausdorff distance. The existence of backsteps implies that
$d_{\rm H}$ cannot be used as the Hausdorff hierarchy's aggregation
index. Indeed, an aggregation index is a positive function $f$
defined on the hierarchy $Y$ satisfying (i) $f(y) =0$ if and only if
$y$ is reduced to a single element of $S$ and (ii) $f(y)<f(y')$ if
$y\in y'$. Equation (\ref{eq:diseq1}) is at variance with condition
(ii). On the other hand, the complete and single hierarchical
algorithm generate a hierarchy indexed through $d_c$ and $d_s$
respectively. Nonetheless, the Hausdorff hierarchy can be indexed
through a proper choice of the aggregation index $f$. This will be
clarified in a forthcoming article. From a more intuitive (physical)
perspective, condition (\ref{eq:diseq1})
can become valid when the sets are rather intertwined, and can be
taken as an indication that, although always mathematically
consistent, the clustering procedure itself at this level of the
dendrogram becomes doubtful, in particular for inherently complex
problems, such as that of clustering stock market companies.

\end{document}